 \documentclass[journal]{IEEEtran}

\usepackage{graphics,color}
\usepackage{amssymb,amsmath,amsthm}
\usepackage{amsfonts,mathrsfs}
\usepackage{makeidx}         
\usepackage{graphicx}        
\usepackage{gensymb}
\usepackage[T1]{fontenc}
\usepackage[utf8]{inputenc}
\usepackage{cite}
\usepackage{tikz}
\usepackage{tkz-tab}
\usepackage{enumitem}

\theoremstyle{definition}\newtheorem*{mysystem}{System}
\theoremstyle{definition}
\theoremstyle{definition}\newtheorem{mydef}{Definition}
\theoremstyle{definition}

\newtheorem{rem}{Remark}
\newtheorem{ex}{Example}

\theoremstyle{plain}
\newtheorem{thm}{Theorem}

\newtheorem{cor}{Corollary}
\newtheorem{prop}[thm]{Proposition}

\newtheorem{theorem}{{Theorem}}
\newtheorem{problem}{{Problem}}

\def\0{{\bf 0}}
\def\1{{\bf 1}}

\def\R{\mathbb{R}}

\def\beq{\begin{equation*}}
\def\eeq{\end{equation*}}
\def\bql{\begin{equation}}
\def\eql{\end{equation}}
\def\bqn{\begin{eqnarray*}}
\def\eqn{\end{eqnarray*}}
\def\bnl{\begin{eqnarray}}
\def\enl{\end{eqnarray}}
\def\bma{\begin{bmatrix}}
\def\ema{\end{bmatrix}}
\def\bmx{\begin{matrix}}
\def\emx{\end{matrix}}
\def\ben{\begin{enumerate}}
\def\een{\end{enumerate}}
\def\bit{\begin{itemize}}
\def\eit{\end{itemize}}
\def\bei{\begin{itemize}}
\def\eei{\end{itemize}}
\def\bet{\begin{tabular}}
\def\eet{\end{tabular}}

\newcommand{\allcaps}[1]{\uppercase\expandafter{#1}}

\begin{document}

\title{Minimal Reachability is {Hard To Approximate}}
\author{A.~Jadbabaie\thanks{{A.~Jadbabaie is with the Institute for Data, Systems, and Society, Massachusetts Institute
of Technology, Cambridge, MA 02139 USA, {\tt jadbabai@mit.edu}.}}, A.~Olshevsky\thanks{{A.~Olshevsky is with the Department of Electrical and Computer Engineering and the Division of Systems Engineering, Boston University, Boston, MA 02215 USA, {\tt alexols@bu.edu}.}}, G.~J.~Pappas\thanks{{G.~J.~Pappas and V.~Tzoumas are with the Department of Electrical and Computer Engineering, University of Pennsylvania, Philadelphia, PA 19104 USA, {\tt pappasg@seas.upenn.edu, vtzoumas@seas.upenn.edu}.}}, V.~Tzoumas}

\maketitle

\begin{abstract}
In this note, we {consider the problem of choosing which nodes of a linear dynamical system should be actuated} so that the state transfer from the system's initial condition to a given {final} state is {possible}. Assuming a standard complexity hypothesis, we show that this problem cannot be efficiently {solved or approximated} in polynomial, {or even quasi-polynomial}, time. \end{abstract}


%
\IEEEpeerreviewmaketitle

\section{Introduction}

During the last decade, researchers in systems, optimization, and control have focused on questions such as:
\begin{itemize} \setlength\itemsep{0.09em}
\item \textit{(Actuator Selection)} How many nodes do we need to actuate in a gene regulatory \mbox{network to control it?~\cite{liu2011controllability, muller2011few}}

\item \textit{(Input Selection)} How many inputs are needed to drive the nodes of a power system to fully control its dynamics?~\cite{zhou2017input}

\item \textit{(Leader Selection)} Which UAVs do we need to choose in a multi-UAV system as leaders for the system to complete a surveillance task despite communication noise?~\cite{clark2014minimizing,clark2014supermodular}
\end{itemize}
The effort to answer such questions has resulted in numerous papers on topics such as actuator placement for controllability~\cite{olshevsky2014minimal, pequito2016framework}; actuator selection and scheduling for bounded control effort~\cite{pasqualetti2014controllability,summers2016submodularity,tzoumas2016minimal, zhao2016scheduling}; resilient actuator placement against failures and attacks~\cite{pequito2017robust,tzoumas2017resilient}; and sensor selection for target tracking and optimal Kalman filtering~\cite{tzoumas2016sensor,tzoumas2016near,zhang2017sensor,carlone2017attention}. In~all these papers the underlying optimization problems have been proven (i)~either polynomially-time solvable~\cite{liu2011controllability, muller2011few,zhou2017input} (ii)~or NP-hard, in which case polynomial-time algorithms have been proposed for their approximate solution~\cite{clark2014minimizing,clark2014supermodular,olshevsky2014minimal, pequito2016framework,pasqualetti2014controllability,summers2016submodularity,tzoumas2016minimal, zhao2016scheduling,pequito2017robust,tzoumas2017resilient,tzoumas2016sensor,tzoumas2016near,zhang2017sensor,carlone2017attention}.

But in systems, optimization, and control, such as in power systems~\cite{amin2008electric,wash-cdc}, transportation networks~\cite{CalTransit}, and  neural circuits~\cite{gu2015controllability,tu2017caveats}, the following problem also arises:

\medskip

\begin{itemize}[leftmargin=*]
\item[] {\bf Minimal Reachability Problem.} 
Given times $t_0$ and $t_1$ such that $t_1>t_0$, vectors $x_0$ and $x_1$, and a linear dynamical system with state vector $x(t)$
such that $x(t_0)=x_0$, find the minimal number of system nodes we need to actuate so that the state transfer from $x(t_0)=x_0$ to $x(t_1)=x_1$ is feasible.
\end{itemize}

\medskip

For example, the stability of power systems is ensured by placing a few generators such that the state transfers from a set of possible initial conditions to the zero state are feasible~\cite{wash-cdc}.  

The minimal reachability problem relaxes the objectives of the applications in~\cite{liu2011controllability, muller2011few,zhou2017input,clark2014minimizing,clark2014supermodular,olshevsky2014minimal, pequito2016framework,pasqualetti2014controllability,summers2016submodularity,tzoumas2016minimal, zhao2016scheduling,pequito2017robust,tzoumas2017resilient,tzoumas2016sensor,tzoumas2016near,zhang2017sensor,carlone2017attention}. For example, in comparison to the actuator placement problem for controllability~\cite{olshevsky2014minimal}, the minimal reachability problem aims to place a few actuators only to make a single transfer between two states feasible, whereas the minimal controllability problem aims to place a few actuators to make the transfer among any two states feasible~\cite{olshevsky2014minimal, pequito2016framework}. 

\begin{figure}[t]
\centering
\begin{tikzpicture}
\tikzstyle{every node}=[draw,shape=circle, minimum size=1.1cm];
\node (v0) at (0:0) {$x_2(t)$};
\node (v1) at (0:1.25) {$x_3(t)$};
\node (v2) at (0:2.5) {$x_4(t)$};
\node[white] (v5) at (0:3.75) {\color{black} $\cdots$};
\node (v3) at (0:5) {$x_{n}(t)$};
\node (v4) at (55:2.5) {$x_1(t)$};

\foreach \from/\to in {v0/v4, v1/v4, v2/v4, v3/v4}
\draw [->] (\from) -- (\to);
\draw
(v0) -- (v4)
(v1) -- (v4)
(v2) -- (v4)
(v3) -- (v4);
\end{tikzpicture}
\caption{{Graphical} representation {of the linear system $\dot{x}_1(t) = \sum_{j=2}^n x_j(t), ~~ \dot{x}_i(t) = 0, ~ i = 2, \ldots, n$};  each node represents an entry of the system's state $(x_1(t),x_2(t),\ldots, x_{n}(t))$, where $t$ represents time; the edges denote that the evolution in time of $x_1$ depends on $(x_2,x_3,\ldots, x_{n})$.}
\label{fig:star}
\end{figure}
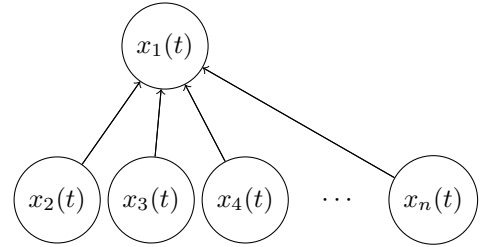

The fact that the minimal reachability problem relaxes the objectives of {the papers}~\cite{liu2011controllability, muller2011few,zhou2017input,clark2014minimizing,clark2014supermodular,olshevsky2014minimal, pequito2016framework,pasqualetti2014controllability,summers2016submodularity,tzoumas2016minimal, zhao2016scheduling,pequito2017robust,tzoumas2017resilient,tzoumas2016sensor,tzoumas2016near,zhang2017sensor,carlone2017attention} is an important distinction whenever we are interested in the feasibility of only a few state transfers {by} a small number of placed actuators.  The reason is that under the objective of minimal reachability the number of placed actuators can be much smaller in comparison to the number of placed actuators under the objective of controllability. For example, in the {system of Fig.~\ref{fig:star}  the number of placed actuators under the objective of minimal reachability from $(0, \ldots, 0)$ to $(1, \ldots, 0)$ is} one, whereas the number of placed actuators under the objective of controllability grows linearly with the system's size.

The minimal reachability problem was introduced in~\cite{tz1}, where it was found to be NP-hard. Similar versions of the reachability problem were studied in the context of power systems in~\cite{wash-cdc} and~\cite{wash-recent}. For~the polynomial-time solution of the reachability problems in~\cite{tz1,wash-cdc,wash-recent}, greedy approximation algorithms were proposed therein. The~approximation performance of these algorithms was {claimed} {by relying} on the modularity result~\cite[Lemma~8.1]{sviridenko2014optimal}, which states that the distance from a point to a subspace created by the span of a set of vectors is {supermodular} in the choice of the vectors.

{In this note, we first show that the modularity result~\cite[Lemma~8.1]{sviridenko2014optimal} is incorrect.  In particular, we show this via a counterexample to~\cite[Lemma~8.1]{sviridenko2014optimal}, and as a result, we prove that the distance from a point to a subspace created by the span of a set of vectors is non-{supermodular} in the choice of the vectors. Then, we~also prove the following strong intractability result for the minimal reachability problem, which is our main contribution in this paper:}

\medskip

\begin{itemize}[leftmargin=*]
\item[] {\bf Contribution~1}{\bf.} 
{Assuming ${\rm NP} \notin {\rm BPTIME}(n^{\rm poly \log \textit{n}})$}, we show that {for each $\delta > 0$}, there is no polynomial-time algorithm that can  distinguish\footnote{{We say that an algorithm can distinguish between two (disjoint) cases $A$~and $B$ if, when fed with an input that is guaranteed to be in either $A$ or~$B$, the algorithm is able to determine which of the two is the case (e.g., by outputing $1$ if the input belongs $A$, and $0$ if it belongs to $B$).}}  between {the two cases where:}
\begin{itemize} \setlength\itemsep{0.09em}
\item the reachability problem has a solution with cardinality~$k$;
\item the reachability problem has no solution with {cardinality $k 2^{\Omega \left( \log^{1-\delta} n \right)}$\!, where $n$ is the dimension of the system.}
\end{itemize} 
\end{itemize}

\medskip

We note that the complexity hypothesis ${\rm NP} \notin{\rm BPTIME}(n^{\rm poly \log \textit{n}})$ means there is no {randomized algorithm which, after running for $O(n^{(\log n)^c})$ time for some constant $c$, outputs correct solutions to problems in ${\rm NP}$ with probability $2/3$; see \cite{AroraBarak} for more details.}

Notably, Contribution~1 remains true even if we allow the algorithm to search for an approximate solution that is relaxed as follows: ~instead of {choosing the actuators to make the state transfer from the initial state $x_0$ to a given final state $x_1$ possible}, some other state~$\widehat{x}_1$ {that satisfies} $\|x_1-\widehat{x}_1\|_2^2\leq \epsilon$ {should be reachable from $x_0$}. {This is a substantial relaxation of the reachability problem's objective, and yet, we show that the intractability result of Contribution~1 still holds}.

The rest of this note is organized as follows. In Section~\ref{sec:min_reach_problem}, we~introduce formally the minimal reachability problem. In~Section~\ref{sec:sub}, we provide a counterexample to~\cite[Lemma~8.1]{sviridenko2014optimal}. In Section~\ref{sec:inapprox}, we~present Contribution~1; in~Section~\ref{app:proof}, we~prove it.
Section~\ref{sec:con} concludes the paper. 

\section{Minimal Reachability Problem} \label{sec:min_reach_problem}

In this section we formalize the minimal reachability problem. We~start by introducing the systems considered in this paper and the notions of system node and of actuated node~set.

\begin{mysystem}
We consider linear systems of the form
\begin{equation}\label{eq:system_model}
\dot{x}(t)=Ax(t)+Bu(t),~~~~~~t \geq t_0,
\end{equation}
where $t_0$ is a given starting time, $x(t) {\in \R^n}$ is the system's state at time $t$, and $u(t) {\in \R^m}$ is the system's input vector. \hfill $\blacktriangleleft$
\end{mysystem}

{In this paper, we want to actuate the minimal number of the nodes of the system in~eq.~\eqref{eq:system_model} to make a desired state-transfer feasible (not achieving necessarily controllability).  We formalize this objective using the following two definitions.
}

\begin{mydef}[System node]
{Given a system as in eq.~\eqref{eq:system_model}, where $x(t)\in \mathbb{R}^n$\!, we let $x_1(t),x_2(t),\ldots, x_n(t)\in \mathbb{R}$ be the components of $x(t)$, i.e., $x(t)=(x_1(t),x_2(t),\ldots, x_n(t))$. We~refer to each $x_i(t)$ as a \emph{system node}. 
\hfill $\blacktriangleleft$}
\end{mydef}
%
%

\begin{mydef}[Actuated node set]
Given a system as in eq.~\eqref{eq:system_model},  we say that the set ${\cal S} \subseteq \{1,2,\ldots, n\}$ is an \emph{actuated node set} if 
the system dynamics can be written as
\begin{equation}\label{eq:input_model}
\dot{x}(t)=Ax(t)+\mathbb{I}(\mathcal{S})Bu(t),~~~~~~t \geq t_0,
\end{equation}
where $\mathbb{I}(\mathcal{S})$ is a diagonal matrix such that if $i \in \mathcal{S}$, the $i$-th entry of $\mathbb{I}(\mathcal{S})$'s diagonal is $1$, otherwise it is $0$. \hfill $\blacktriangleleft$
\end{mydef}


The definition of $\mathbb{I}(\cal S)$ in eq.~\eqref{eq:input_model} implies that the input $u(t)$ affects only the system nodes $x_i(t)$ where $i \in \mathcal{S}$. 

\begin{problem}[Minimal Reachability] \label{pr:minimal_reachability}
Given
\begin{itemize} \setlength\itemsep{0.09em}
\item times $t_0$ and $t_1$ such that $t_1>t_0$,
\item {vectors $x_0,~x_1 \in \R^n$}\!\!, and
\item a system $\dot{x}(t)=Ax(t)+Bu(t),~t\geq t_0,$ as in eq.~\eqref{eq:system_model}, with initial condition $x(t_0)=x_0$,
\end{itemize}
find an actuated node set with minimal cardinality such that there exists an input ${u(t)}$ defined over the time interval $(t_0,t_1)$ that achieves $x(t_1)=x_1$. \mbox{Formally}, using the notation $|\mathcal{S}|$ to denote the cardinality of a set $\mathcal{S}$:
\[ \underset{\mathcal{S} \subseteq \{1,2,\ldots,n\}}{\text{minimize}} 
 \;  \quad  |\mathcal{S}| \] such that there exist $u: (t_0,t_1) \mapsto \mathbb{R}^m\!\!,~ x: (t_0,t_1) \mapsto \mathbb{R}^n$ with 
\begin{eqnarray*} 
  &\dot{x}(t)=Ax(t)+\mathbb{I}(\mathcal{S})Bu(t), &~~~~~~t \geq t_0, \\ 
 &x(t_0)=x_0,~~~~~x(t_1)=x_1.   &
\end{eqnarray*}
\end{problem}

{A special case of interest is when $B$ is the identity matrix. Then} minimal reachability
asks for the fewest system nodes to be actuated {directly} so that at time~$t_1$ the state $x_1$ is reachable from the system's initial condition~$x(t_0)=x_0$.

\section{{Non-{super}modularity of distance from point to subspace}}\label{sec:sub} 

{In this section, we provide a counterexample to the {super}modularity result~\cite[Lemma~8.1]{sviridenko2014optimal}.
{We begin with some notation.} In~particular, 
given a matrix $M \in \R^{n \times n}$\!\!, a vector $v \in \R^n$\!\!, and a set $\mathcal{S}\subset \{1, \ldots, n\}$, let $M(\mathcal{S})$ denote the matrix obtained by throwing away columns of $M$ not in $\mathcal{S}$.
In addition, for any set $\mathcal{S}\subset \{1, \ldots, n\}$, we define the set function
\[ f(\mathcal{S}) = {\rm dist}^2(v,{\rm Range}( M(\mathcal{S})) ), \] where ${\rm dist}(y,X)$ is the distance from a point to a subspace:
\begin{equation*}
{\rm dist}(y,X) = \min_{x \in X} ||y-x||_2.
\end{equation*}

We show there exist a vector $v$ and a matrix $M$ such that the function $f:\{1,2,\ldots,n\}\mapsto {\rm dist}^2(v,{\rm Range}( M(\mathcal{S})) )$ is non-{super}modular.  We start by defining the monotonicity and {super}modularity of set functions.
%

\begin{mydef}[Monotonicity]
Consider any finite  set $\mathcal{V}$. The set function $f:2^\mathcal{V}\mapsto \mathbb{R}$ is non-decreasing if and only if for any $\mathcal{A}\subseteq \mathcal{A}'\subseteq\mathcal{V}$, {we have} $f(\mathcal{A})\leq f(\mathcal{A}')$.\hfill $\blacktriangleleft$
\end{mydef}
In words, a set function $f:2^\mathcal{V}\mapsto \mathbb{R}$ is non-decreasing if and only if by adding elements in any set $\mathcal{A}\subseteq\mathcal{V}$ we cannot decrease the value of~$f(\mathcal{A})$.
 
\begin{mydef}[{Super}modularity~{\cite[Proposition 2.1]{nemhauser78analysis}}]\label{def:sub}
Consider any finite set $\mathcal{V}$.  The set function $f:2^\mathcal{V}\mapsto \mathbb{R}$ is {super}modular if and only if for any $\mathcal{A}\subseteq \mathcal{A}'\subseteq\mathcal{V}$ and $x\in \mathcal{V}$, \belowdisplayskip=-12pt{$$f(\mathcal{A})-f(\mathcal{A}\cup \{x\})\geq f(\mathcal{A}') - f(\mathcal{A}'\cup \{x\}).$$} \hfill $\blacktriangleleft$
\end{mydef}
In words, a set function $f:2^\mathcal{V}\mapsto \mathbb{R}$ is {super}modular if and only if it satisfies the following diminishing returns property: for any element $x\in \mathcal{V}$, the {marginal decrease $f(\mathcal{A})-f(\mathcal{A}\cup \{x\})$} diminishes as the set $\mathcal{A}$ grows; equivalently, for any $\mathcal{A}\subseteq \mathcal{V}$ and $x\in \mathcal{V}$, {$f(\mathcal{A})-f(\mathcal{A}\cup \{x\})$} is non-increasing.

\begin{ex}\label{ex:nonsuper} We show that for
\[ v = \left( \begin{array}{c} -1 \\ 1 \\ 1 \end{array} \right), ~~~~~~ M = 
\left( \begin{array}{ccc} 1  &  0  & 1 \\
       1 & 1 & 0 \\
       0 & 0 & 1
       \end{array} \right)  , \]
the set function $f:2^{\{1,2,3\}}\mapsto {\rm dist}^2(v,{\rm Range}( M(\mathcal{S})) )$ is non-{super}modular.  In particular, since the vector $v$ is orthogonal to the first and third columns of $M$,
\begin{eqnarray*}  f(\{1\})={\rm dist}^2(v,M(\{1\})) & = & ||v||_2^2  \\
f(\{1,3\})={\rm dist}^2(v,M(\{1,3\})) & = & ||v||_2^2 
\end{eqnarray*}
Therefore,
 \[ f(\{1\}) - f(\{1,3\}) = 0. \] 
At the same time, the span of the first two columns of $M$ is the subspace $\{ x \in \R^3: x_3 = 0\}$. Thus,
\[f(\{1,2\})= {\rm dist}^2(v,M(\{1,2\})) = 1. \]
Also, since the three  columns of $M$ are linearly independent,
\[  f(\{1,2,3\})={\rm dist}^2(v,M(\{1,2,3\})) = 0, \]
and as a result,
\[ f(\{1,2\}) - f(\{1,2,3\}) =1. \]
In sum,
\[ f(\{1,2\}) - f(\{1,2,3\})>f(\{1\}) - f(\{1,3\}); \]
hence, for the vector $v$ and matrix $M$ in this example, $f:2^{\{1,2,3\}}\mapsto {\rm dist}^2(v,{\rm Range}( M(\mathcal{S})) )$ is non-{supermodular}.
\hfill $\blacktriangleleft$
\end{ex}

{We remark that the same argument as in Example~\ref{ex:nonsuper} shows that the set function $g:\{1,2,\ldots,n\}\mapsto \mathbb{R}$ such that $g({\cal S})= [{\rm dist}(v,{\rm Range}( M(\mathcal{S}))]^c$ is not {supermodular} for any $c>0$.}

\section{Inapproximability of Minimal Reachability Problem}\label{sec:inapprox}

We show that,  {subject to a widely believed conjecture in complexity theory}, there is no efficient algorithm that solves, even approximately, Problem~\ref{pr:minimal_reachability}. Towards the statement of this result, we {next introduce} a definition of approximability and the definition of quasi-polynomial running time. 

\begin{mydef}[Approximability]\label{def:approx}
Consider the minimal reachability Problem~\ref{pr:minimal_reachability}, and let the set $\mathcal{S}^\star$ to denote one of its optimal solutions. {We say that an algorithm renders} Problem~\ref{pr:minimal_reachability}  \emph{$(\Delta_1(n), \Delta_2(n))$-approximable} if it returns a set $\mathcal{S}$ such~that:
\begin{itemize}  \setlength\itemsep{0.09em}
\item  there is a state $\widehat{x}_1$ such that there is an input $u(t)$ such that at time $t_1$ we have  $x(t_1)=\widehat{x}_1$ and $||\widehat{x}_1 - x_1\|_2\leq \Delta_1(n)$;
\item the cardinality of the set $\mathcal{S}$ is at most $\Delta_2(n)|\mathcal{S}^\star|$. \hfill $\blacktriangleleft$
\end{itemize}
\end{mydef}

{Hence, the definition of $(\Delta_1(n), \Delta_2(n))$-approximability allows some slack both in the quality of the reachability requirement (first point in the itemization in Definition~\ref{def:approx}), and in the number of actuators utilized to achieve it (second point in the itemization in Definition~\ref{def:approx}).}

{We introduce next the definition of quasi-polynomial algorithms, using the following big O notation.}

{\begin{mydef}[Big O notation]\label{def:bigO} Let $\mathbb{N}$ be the set of natural numbers, and consider two functions $h:\mathbb{N}\mapsto \mathbb{R}$ and $g:\mathbb{N}\mapsto \mathbb{R}$ that take only non-negative values. The \emph{big O notation} in the equality $h(n)=O(g(n))$ means there exists some constant $c>0$ such that for all large enough $n$, it is $h(n)\leq c g(n)$. \hfill $\blacktriangleleft$
\end{mydef}}

{Definition~\ref{def:bigO}, given a non-negative function $g$, implies that $O(g(n))$ denotes the collection of non-negative functions $h$ that are bounded asymptotically by $g$, up to a constant factor.}

\begin{mydef}[Quasi-polynomial running time]\label{def:quasi}
An algorithm is \emph{quasi-polynomial} if it runs in $2^{O\left[( \log n)^c \right]}$ time, where $c$ is a constant. 
\hfill $\blacktriangleleft$
\end{mydef}

{We note that any polynomial-time algorithm is  a quasi-polynomial time algorithm since $n^k = 2^{k \log n}$\!. At the same time, a quasi-polynomial algorithm is asymptotically faster than an exponential-time algorithm, since exponential-time algorithms run in $O(2^{n^\epsilon})$ time, for some $\epsilon>0$. }


{\begin{mydef}[Big Omega notation]\label{def:bigOmega} Let $\mathbb{N}$ be the set of natural numbers, and consider the functions $h:\mathbb{N}\mapsto \mathbb{R}$ and $g:\mathbb{N}\mapsto \mathbb{R}$ that take only non-negative values. The \emph{big Omega notation} in the equality $h(n)=\Omega(g(n))$ means that there exists some constant $c>0$ such that for all large enough $n$, it is $h(n)\geq c g(n)$. \hfill $\blacktriangleleft$
\end{mydef}}

{Definition~\ref{def:bigOmega}, given a non-negative $g$, implies that $\Omega(g(n))$ denotes the collection of non-negative functions $h$ that are lower bounded asymptotically  by $g$, up to a constant~factor.}

{We present next {our main result in this paper}.

\smallskip

\begin{theorem}[Inapproximability] \label{mainthm} {For each $\delta \in (0,1)$,} there is a {collection of instances} of Problem~\ref{pr:minimal_reachability} where:
\begin{itemize} \setlength\itemsep{0.09em}
\item the initial condition is $x(t_0)=0$;
\item the final state $x_1$ is of the form $[1,1,\ldots,1,0, 0, \ldots, 0]^\top$\!;
\item the input matrix is $B=I$, where $I$ is the identity matrix,
\end{itemize} 
 {along with} {a polynomial $\Delta_1(n)$ and a function
$\Delta_2(n)=2^{\Omega(\log^{1-\delta} n )}$\!, {such that unless ${\rm NP} {\in}$ ${\rm BPTIME(\textit{n}^{\rm poly \log \textit{n}})}$, there is no quasi-polynomial algorithm rendering Problem~\ref{pr:minimal_reachability}  $(\Delta_1(n), \Delta_2(n))$-approximable.}} 
\end{theorem} 

\smallskip

Theorem~\ref{mainthm} says that {if} ${\rm NP} \notin {\rm BPTIME(\textit{n}^{\rm poly \log \textit{n}})}$ there is no polynomial time algorithm {(or quasi-polynomial time algorithm)} that can {choose which entries of the system's $x$ state to actuate} so that $x(t_1)$ is even approximately close to a desired state $x_1=[1,1,\ldots,1,0, 0, \ldots, 0]^\top$ at time $t_1$.  

To make sense of Theorem~\ref{mainthm}, {first observe that we can always actuate every entry of the system's state, i.e., we can choose $\mathcal{S}=\{1,2,\ldots,n\}$. This means every system is $(0,n)$-approximable; let us rephrase this by saying that every system is $(0,2^{\log n})$ approximate.  Theorem~\ref{mainthm} tells us that we cannot achieve $(0, 2^{{O} (\log^{1-\delta} n )})$-approximability {\em for any $\delta > 0$}. In other words, improving the guarantee of the strategy that actuates every state by just a little bit, in the sense of replacing $\delta = 0$ with some $\delta > 0$, is not possible ---subject to the complexity-theoretic hypothesis ${\rm NP} \notin {\rm BPTIME(\textit{n}^{\rm poly \log \textit{n}})}$.  Furthermore, the theorem tells us it remains impossible even if we allow ourselves some error $\Delta(n)$ in the target state, i.e., even $(\Delta(n), 2^{{O} (\log^{1-\delta} n )})$--approximability is ruled out.}

\begin{rem}
In~\cite[Theorem~3]{tz1} it is claimed that for any $\epsilon > 0$ the minimal reachability Problem~\ref{pr:minimal_reachability} is $\left( \epsilon, O \left( \log \frac{n}{\epsilon} \right) \right)$-approximable, which contradicts Theorem~\ref{mainthm}. However, the proof of this claim was based on~\cite[Lemma 8.1]{sviridenko2014optimal}, which we proved incorrect in Section~\ref{sec:sub}. 
\hfill $\blacktriangleleft$
\end{rem}

\begin{rem} {The minimal controllability problem~\cite{olshevsky2014minimal} seeks to place the fewest number of actuators to make the system controllable. Theorem~\ref{mainthm} is arguably surprising, as it was shown in \cite{olshevsky2014minimal} that the sparsest set of actuators for controllability can be approximated to a multiplicative factor of $O(\log n)$ in polynomial time. By contrast, we showed in this note that an almost exponentially worse approximation ratio {\em cannot} be achieved for minimum reachability.  }
\hfill $\blacktriangleleft$
\end{rem}

\section{Proof of Inapproximability of Minimal Reachability}\label{app:proof}

{We next provide a proof of our main result, namely Theorem~\ref{mainthm}. We use some standard notation throughout: ${\bf 1}_k$ is the all-ones vector in $\R^k$, ${\bf 0}_k$ is the zero vector in $\R^k$, and $e_k$ is the $k$'th standard basis vector. We begin with some standard definitions related to the reachability space of a linear system.}

\setcounter{subsection}{0}
 
\subsection{Reachability Space for continuous-time linear systems}

\begin{mydef}[Reachability space]\label{def:reach}
Consider a system $\dot{x}(t)=Ax(t)+Bu(t)$ as in eq.~\eqref{eq:system_model} whose size is $n$.
The ${\rm Range} ([B,~ AB,~ A^2 B, \ldots, ~A^{n-1}B])$ is called the \emph{reachability space} of  $\dot{x}(t)=Ax(t)+Bu(t)$. \hfill $\blacktriangleleft$
\end{mydef}

The reason why {Definition~\ref{def:reach} is called the reachability space is explained in the following proposition}.

\begin{prop}[{~\hspace{-1mm}\cite[Proof of Theorem~6.1]{chi-TsongChen}}]
\label{prop:reach_subspace}
Consider a system as in eq.~\eqref{eq:system_model},
with initial condition $x_0$. 
There exists a real input $u(t)$ defined over the time interval $(t_0,t_1)$ such that {the solution of} $\dot{x}=Ax+Bu,~~{x(t_0)=x_0}$ {satisfies} $x(t_1)=x_1$ if and only if
\begin{equation*}
 x_1-e^{A(t_1-t_0)}x_0 \in {\rm Range} ([B,~ AB,~ A^2 B, \ldots, ~A^{n-1}B]).
\end{equation*}
\end{prop}

\medskip

The notion of reachability space {allows us to redefine} the minimal reachability Problem~\ref{pr:minimal_reachability} {as follows}.

\begin{cor}\label{cor:reachability}
Problem~\ref{pr:minimal_reachability} is equivalent~to
\begin{equation*}
\begin{aligned}
 \underset{\mathcal{S} \subseteq \{1,2,\ldots,n\}}{\text{minimize}} 
 \;  & |\mathcal{S}| \\
\text{such that \hspace{1mm}} 
 &  x_1-e^{A(t_1-t_0)}x_0 \in \\
 & {\rm Range} ([\mathbb{I}(\mathcal{S})B,~ A\mathbb{I}(\mathcal{S})B, \ldots, ~A^{n-1}\mathbb{I}(\mathcal{S})B]).
\end{aligned}
\end{equation*}
\end{cor}

Overall, Problem~\ref{pr:minimal_reachability} is equivalent to picking the fewest rows of the input matrix~$B$ such that $x_1-e^{A(t_1-t_0)}x_0$ is in the {linear span of the columns of $[\mathbb{I}(\mathcal{S})B,~ A\mathbb{I}(\mathcal{S})B,~ A^2 \mathbb{I}(\mathcal{S})B, \ldots, ~A^{n-1}\mathbb{I}(\mathcal{S})B]$}.

\subsection{Variable Selection Problem}\label{sub:variable_selection}

{We show the intractability of the minimum reachability by reducing it to the {\em variable selection} problem, defined next. }

\begin{problem}[Variable Selection]\label{pr:selection}
{Let $U \in \R^{m \times l}$, $z \in \R^m$, and let $\Delta$ be a positive number. The variable selection problem is} to pick {$y \in \R^l$ that is an optimal solution to the following optimization problem.}
\begin{equation*}
\begin{aligned}
 \underset{y \in \mathbb{R}^l}{\text{minimize}} 
 \;  \quad & \|y\|_0 \\
\text{such that\hspace{1mm}} 
 \quad & \|Uy-z\|_2\leq \Delta,\\
\end{aligned}
\end{equation*} {where $||y||_0$ refers to the number of non-zero entries of $y$.}
\end{problem}

The variable selection Problem~\ref{pr:selection} is found in~\cite{varsel} to be inapproximable, even in quasi-polynomial time:

\begin{theorem}[\hspace{-0.15mm}{\cite[{Proposition 6}]{varsel}}]\label{th:inapprox_selection}
Unless it is ${\rm NP} {\in}$ ${\rm BPTIME(\textit{n}^{\rm poly \log \textit{n}})}$, for each $\delta \in (0,1)$ there exist:
\begin{itemize} \setlength\itemsep{0.09em}
\item {a function $q_1(l)$ which is in {$2^{\Omega(\log^{1-\delta} l)}$}; }
\item {a polynomial $p_1(l)$ which is in $O(l)$;}\footnote{{In this context, a function with a fractional exponent is considered to be a polynomial, e.g., $l^{1/5}$ is considered to be a polynomial in $l$.}}
\item {a polynomial $\Delta(l)$;}
\item {a polynomial $m(l)$,} \end{itemize} 
{and a zero-one} $m(l) \times l$ matrix $U$ such that no quasi-polynomial time algorithm can distinguish between the following two cases {for large $l$}: 
\begin{enumerate} \setlength\itemsep{0.09em}
\item There exists a vector $y \in {\mathbb{R}^l}$ such that $Uy = \1_{m(l)}$ and $||y||_{0} \leq p_1(l)$. 
\item For any vector $y \in \R^l$ such that $||Uy - \1_{m(l)}||_2^2 \leq \Delta(l)$, we have $||y||_0 \geq p_1(l) q_1(l)$.
\end{enumerate}
\end{theorem}  

{Informally, unless ${\rm NP} {\in} {\rm BPTIME(\textit{n}^{\rm poly \log \textit{n}})}$, Theorem~2 says that Problem~2 is inapproximable even in quasi-polynomial time, in the sense that  for large $l$ there is no quasi-polynomial algorithm that can distinguish between the two mutually exclusive cases 1) and 2). To see that these cases are indeed mutually exclusive for large $l$, observe that $q_1(l) > 1$ when $l$ is large, because $q_1(l) = 2^{ \Omega \left( \log^{1-\delta} l \right)}$\!.}

\subsection{Sketch of Proof of Theorem~\ref{mainthm}}

We begin by sketching the intuition behind the proof of Theorem~\ref{mainthm}.  {Our general approach is to} find instances of Problem~\ref{pr:minimal_reachability} that are as hard as inapproximable instances of the variable selection Problem~\ref{pr:selection}. {We begin by discussing a construction that does {\em not} work, and then explain how to fix~it.}

Given the matrix $U$ {coming from a} variable selection Problem~\ref{pr:selection}, we first {attempt to} construct an instance of the minimal reachability Problem~\ref{pr:minimal_reachability} where:
\begin{itemize} \setlength\itemsep{0.09em}
\item the system's initial condition is $x(t_0)=0$;
\item the destination state $x_1$ at time $t_1$ is of the form $[\1,\0]^\top$ (the exact dimensions of $\1$ and $\0$ are to be determined);
\item the input matrix is $B=I$;
\item the system matrix $A$ is
\begin{equation}\label{eq:A_not}
A = \left( \begin{array}{cc} 0 & U \\ 0 & 0 \end{array} \right),
\end{equation}
{where the number of zeros is large{ so that $A^2=0$}.} 
\end{itemize} 

{Whereas the variable selection problem involves finding the smallest set of columns of $U$ so that a certain vector is in their span, for the minimum reachability problem, every time we add the $k$-th state to the set of actuated variables $\cal S$, the reachability span expands by adding the span of the set of columns of the controllability matrix that correspond to the vector~$e_k$ being added in $\mathbb{I}(\cal S)$. In particular, for the above construction, because $A^2=0$, when the $k$-th state is added to the set of actuated variables, the span of the two columns $e_k$ and $Ue_{k}$ is added to the reachability space.}

{In other words, with the above construction we are basically constrained to make ``moves'' which add columns in pairs, and we are looking for the smallest number of such ``moves'' making a certain vector lie in the span of the columns. It~should be clear that there is a strong parallel between this and variable selection (where the columns are added one at a time). However, because the columns are being added in pairs, this attempt to connect minimum reachability with variable selection does not quite work.}
{To fix this idea, we want only the columns of $U$ to contribute meaningfully to the addition of the span, with any vectors $e_k$ we add along the way being redundant; this would reduce minimal reachability to exactly variable selection.  We accomplish this by further defining:}
\[ U' = \left( \begin{array}{c} U \\ U \\ \vdots \\ U \end{array} \right), \] where we stack $U$ some large number of times  (to be determined in the main proof of Theorem~\ref{mainthm} at Section~\ref{subsec:proof}). We~then set:
\begin{equation}\label{eq:A_yes}
 A = \left( \begin{array}{cc} 0 & U' \\ 0 & 0 \end{array} \right).
\end{equation}
{The idea is because $U$ is stacked many times, adding a column of $U$ to a set of vectors expands the span much more than adding any vector $e_k$, so there is never an incentive to consider the contributions of any $e_k$ to the reachability space.} 

{We make the aforementioned construction of the system matrix $A$ precise: {given a matrix $M \in \R^{m \times l}$\!, for $n \geq \max\{m,l\}d$} we define $\phi_{n,d}(M)$ to be the $n \times n$ matrix which stacks{$M$} in the top-right hand corner $d$ times. For example, 
\[ M = \left( \begin{array}{cc} 1 & 2 \\ 3 & 4 \end{array} \right), ~~~ \phi_{5,2}(M) =
\left( \begin{array}{ccccc}
0 & 0 & 0 & 1 & 2 \\
0 & 0 & 0 & 3 & 4 \\ 
0 & 0 & 0 & 1 & 2 \\
0 & 0 & 0 & 3 & 4 \\
0 & 0 & 0 & 0 & 0
\end{array}
\right)
,\] i.e., $\phi_{5,2}(M)$ stacks $M$ twice, and then pads it with enough zeros to make the resulting matrix $5 \times 5$. {Observe that $\phi_{n,d}(M)^2 = 0$ for $n \geq  \max\{m,l\}(d+1)$.  Overall, in the next section, we set $A=\phi_{n,d}(U)$ for large enough $d$, and $n =  \max\{m,l\}(d+1)$, and we prove Theorem~1}. }

\subsection{Proof of Theorem~\ref{mainthm}}\label{subsec:proof}

{Adopting the notation in Theorem~2, we focus on problem instances where for large enough $l$ it is $q_1(l)>1$, per the proof of Theorem~2, i.e., of~\cite[Theorem~2]{varsel}.  In addition, we let $d = \lceil p_1(l) q_1(l) \rceil$, and $n = \max \{m(l),l\}(d+1)$. Moreover, for simplicity, we use henceforth $m$ and $m(l)$ interchangeably. 
{Finally, we consider the instances of Problem~1 where:}
\begin{itemize} \setlength\itemsep{0.09em}
\item the initial condition is $x(t_0)={\mathbf{0}_n}$;
\item the destination state $x_1$ at time $t_1$ is $[\1_{md}^\top, \0_{n-md}^\top]^\top$\!;
\item the input matrix is $B=I$, where $I$ is the identity matrix;
\item the system matrix {is $A = \phi_{n,d}(U)$.} 
\end{itemize} 
}
{Given the above, to prove Theorem~\ref{mainthm} we first define the following four statements: 
\begin{enumerate} \setlength\itemsep{0.09em}
\item[S1)] There exists a vector $y\in \R^l$ such that $Uy = \1_{m}$ and $||y||_0 \leq p_1(l)$.
\item[S2)] For any vector $y \in \R^l$ such that $||Uy-\1_m||_2^2 \leq \Delta(l)$, we have $||y||_0 \geq p_1(l) q_1(l)$.
\item[S1$'$)] There exists a set $\mathcal{S}\subseteq \{1,2,\ldots,n\}$ with $|\mathcal{S}|\leq p_1(l)$ such that the state $x_1=[\1_{md}^\top, \0_{n-md}^\top]^\top$ is reachable.
\item[S2$'$)] There is no set $\mathcal{S}\subseteq \{1,2,\ldots,n\}$ with cardinality strictly less than $p_1(l) q_1(l)$ that makes reachable some $\widehat{x}_1$ with $||\widehat{x}_1 - [\1_{md}^\top, \0_{n-md}^\top]^\top||_2^2 \leq \Delta(l)$.
\end{enumerate}
Recall that in Section~\ref{sub:variable_selection} we stated that the statements S1-S2 are mutually exclusive for $q_1(l)>1$ (which is the case for the instances we consider in this proof), and that Theorem~2 implies there is no quasi-polynomial algorithm (unless ${\rm NP} {\in}$ ${\rm BPTIME(\textit{n}^{\rm poly \log \textit{n}})}$) that can distinguish between S1 and S2.}  

{Given the above, we next proceed with the proof of Theorem~\ref{mainthm} by proving first that statement S1 implies statement~S1$'$, and then that also statement S2 implies statement S2$'$\!.  
}

{\paragraph*{Proof that statement S1 implies statement S1$'$} We prove that if statement S1 is true, then statement S1$'$ also is.}  In particular, suppose there exists a vector $y\in \R^l$ with $Uy = \1_{m}$ and $||y||_0 \leq p_1(l)$ (statement S1). In~this case, we claim there exists a set $\mathbb{S}\subseteq \{1,2,\ldots,n\}$ with $|\mathbb{S}|\leq p_1(l)$ such that ${x_1=}[\1_{md}^\top, \0_{n-md}^\top]^\top$ is reachable (statement S1$'$). Indeed, let $S$ be a set of columns of $U$ that have $\1_m$ in their span, and set $\mathbb{S}=\{k+n-l~|~k \in S\}$. Then $|\mathbb{S}|\leq p_1(l)$, and
\begin{equation}\label{eq:aux-1} 
\1_m = \sum_{k \in S} y_kU_k, 
\end{equation}
where $y_k$ denotes the $k$-th element of the vector $y$, and $U_k$ denotes the $k$-th column of the matrix $U$. Due to eq.~\eqref{eq:aux-1}, {we can rewrite the vector ${x_1=}[\1_{md}^\top, \0_{n-md}^\top]^\top$ as follows:}
 \begin{align}  \left( \begin{array}{c} \1_{md} \\ \0_{n-md} \end{array} \right) =   \left( \begin{array}{c} \1_m \\ \1_m  \\ \vdots \\ \1_m  \\ \0_{n-md} \end{array} \right)  &=  \sum_{k \in S} \left( \begin{array}{c} y_kU_k \\ y_kU_k \\ \vdots \\ y_kU_k \\ \0_{n-md} \end{array} \right)\nonumber \\
 &= \sum_{k \in S} y_kA_{k+n-l}, \label{eq:aux-3}
 \end{align}
{{where the vector $\1_m$ in the second term from the left is repeated $ \lceil p_1(l) q_1(l) \rceil$ times, since $d=\lceil p_1(l) q_1(l) \rceil$}, and where the final step (eq.~\eqref{eq:aux-3}) follows by definitions of $A$ as $A=\phi_{n,d}(U)$,
and where $A_{k+n-l}$ denotes the $(k+n-l)$-th column of $A$. Now,}  each of the vectors $A_{k+n-l}$ in the last term is a column of $ A\mathbb{I}(\mathbb{S})$, so $[\1_{md}^\top, \0_{n-md}^\top]^\top$ indeed lies in the range {of the controllability matrix} and, as a result, the state $x_1=[\1_{md}^\top, \0_{n-md}^\top]^\top$ is reachable  by actuating $\mathbb{S}$.
}
}

{\paragraph*{Proof that statement S2 implies statement S2$'$} We prove that if the statement S2 is true, then the statement S2$'$ also is.}  In particular, per statement S2 suppose that any vector $y$ with $||Uy-\1_m||_2^2 \leq \Delta(l)$ has the property that $||y||_0 \geq p_1(l) q_1(l)$. We claim that in this case there is no set $\mathbb{S}\subseteq \{1,2,\ldots,n\}$ with cardinality strictly less than $p_1(l) q_1(l)$ that makes reachable some $\widehat{x}_1$ with $||\widehat{x}_1 - [\1_{md}^\top, \0_{n-md}^\top]^\top||_2^2 \leq  \Delta(l)$  (statement S2$'$). 
To prove this, assume the contrary, i.e., {assume there exists $\mathbb{S}$ with cardinality strictly less than $p_1(l) q_1(l)$ that makes reachable some $\widehat{x}_1$ with $||\widehat{x}_1 - [\1_{md}^\top, \0_{n-md}^\top]^\top||_2^2 \leq \Delta(l)$ ---}we call this assumption~A1. We obtain a contradiction {as follows}:
the pigeonhole principle implies that in the set $\{1,2,\ldots,md\}$ there is some interval $\mathbb{E} = \{\kappa m+1,  \kappa m+2, \ldots, \kappa m+m\}$, where $\kappa$ is a non-negative integer, such that  $\mathbb{S} \cap \mathbb{E} = \emptyset$, because $|\mathbb{S}|<p_1(l) q_1(l)$ and $md\geq m {\lceil} p_1(l) q_1(l) {\rceil}$. 
 Define the vector $\widehat{x}_\mathbb{E} \in \R^m$ by taking the rows of $\widehat{x}_1$ corresponding to indexes in $\mathbb{E}$. {Then, $$||[\widehat{x}_\mathbb{E}  - \1_m||_2^2 \leq \Delta(l),$$ since $\widehat{x}_1$ with $||\widehat{x}_1 - [\1_{md}^\top, \0_{n-md}^\top]^\top||_2^2 \leq \Delta(l)$. 
Moreover, 
we next prove that $\widehat{x}_\mathbb{E}$ is in the span of $|\mathbb{S}|$ columns of $U$.  To this end, we make the following observations:
 since $\widehat{x}_1$ is reachable, it is:
  \begin{align}\label{eq:aux_10}
 \widehat{x}_1 \in &\;{\rm Range}[\mathbb{I}(\mathbb{S}), A\mathbb{I}(\mathbb{S}), A^2\mathbb{I}(\mathbb{S}),\ldots, A^{n-1}\mathbb{I}(\mathbb{S})]=\nonumber\\
& \;{\rm Range}[\mathbb{I}(\mathbb{S}), A\mathbb{I}(\mathbb{S})],
 \end{align}
where the equality in eq,~\eqref{eq:aux_10} holds since {$A^2=0$}. Now, eq.~\eqref{eq:aux_10} implies there exists a vector $z$ such that:
  \begin{equation}\label{eq:aux_11}
[\mathbb{I}(\mathbb{S}), A\mathbb{I}(\mathbb{S})]z= \widehat{x}_1.
 \end{equation}
If we break up the set $\mathbb{S}$ into two sets, (i)~the set of indexes corresponding to $A$'s first $n-l$ columns, which we denote henceforth by $\mathbb{S}_{1:n-l}$, and (ii)~the set of indexes corresponding to $A$'s last $l$ columns, which we denote henceforth by $\mathbb{S}_{n-l+1:n}$, such that $\mathbb{S} =\mathbb{S}_{1:n-l} \cup \mathbb{S}_{n-l+1:n}$, and recall $A$'s definition, we can write the term $A\mathbb{I}(\mathbb{S})$ in eq.~\eqref{eq:aux_11} as follows:
\begin{align}
A\mathbb{I}(\mathbb{S})&=\left(\begin{array}{cc} 0 & U'\\ 0 & 0 \end{array}\right) \left(\begin{array}{cc} \mathbb{I}(\mathbb{S}_{1:n-l}) & 0\\ 0 & \mathbb{I}(\mathbb{S}_{n-l+1:n})\end{array}\right)\nonumber\\
&= \left(\begin{array}{cc}0 & U'\mathbb{I}(\mathbb{S}_{n-l+1:n})\\ 0 & 0\end{array}\right),\label{eq:aux_12}
\end{align} 
where $U'$ is, per the definition of $A$, the matrix that is created by stacking $d$ copies of $U$ the one on top of the other.  Therefore, using this definition of $U'$\!, the term $U'\mathbb{I}(\mathbb{S}_{n-l+1:n})$ in eq.~\eqref{eq:aux_12} is re-written as follows:
\begin{align}
U'\mathbb{I}(\mathbb{S}_{n-l+1:n})&=\left(\begin{array}{c}  U\mathbb{I}(\mathbb{S}_{n-l+1:n})\\ U\mathbb{I}(\mathbb{S}_{n-l+1:n})\\ \vdots \\ U\mathbb{I}(\mathbb{S}_{n-l+1:n}) \end{array}\right),\label{eq:aux_13}
\end{align}
where the term $U\mathbb{I}(\mathbb{S}_{n-l+1:n})$ is repeated $d$ times.  Let now~$z_1$ be the vector that is constructed by $z$ by keeping all the elements of $z$ that in eq.~\eqref{eq:aux_11} multiply the matrix $\mathbb{I}({\mathbb{S}})$, and let~$z_2$ be the vector that is constructed by $z$ by keeping all the elements of $z$ that in eq.~\eqref{eq:aux_11} multiply the non-zero part of the matrix $A\mathbb{I}(\mathbb{S})$, which is stated in eq.~\eqref{eq:aux_13}.
Then, due to eq.~\eqref{eq:aux_12} and eq.~\eqref{eq:aux_13}, the eq.~\eqref{eq:aux_11} gives:
\begin{equation}\label{eq:aux_14}
\mathbb{I}(\mathbb{S})z_1+\left(\begin{array}{c}  U\mathbb{I}(\mathbb{S}_{n-l+1:n})z_2\\ U\mathbb{I}(\mathbb{S}_{n-l+1:n})z_2\\ \vdots \\ U\mathbb{I}(\mathbb{S}_{n-l+1:n})z_2 \\ 0 \end{array}\right)= \widehat{x}_1,
\end{equation}
Moreover, $\widehat{x}_\mathbb{E}$, due to its definition, is in the span of the vectors obtained by taking the rows $\kappa m + 1, \ldots, \kappa m + m$ of the columns of the reachability matrix $[\mathbb{I}(\mathbb{S}), A\mathbb{I}(\mathbb{S})]$; in particular, since it is $\mathbb{S} \cap \mathbb{E}  = \emptyset$, from eq.~\eqref{eq:aux_14} we get:
\begin{align}\label{eq:aux_40} 
U\mathbb{I}(\mathbb{S}_{n-l+1:n})z_2=\widehat{x}_\mathbb{E},
\end{align}
and indeed we have shown that the vector $\widehat{x}_\mathbb{E}$ 
is in the span of at most $|\mathbb{S}|$ columns of $U$ (eq.~\eqref{eq:aux_40}).} {The contradiction is now obtained because assumption A1 tells us that $|\mathbb{S}| < p_1(l) q_1(l)$ while the statement S2 (which we have assumed initially to hold) tells us the opposite. As a result, the truth of statement S2 implies the truth of statement S2$'$\!.} 
 
In sum, we proved that the statement S1 implies the statement S1$'$, as well as, that the statement S2 implies the statement S2$'$ and, as a result, we showed how Problem~1 can be reduced to the (inapproximable in quasi-polynomial time) Problem~2.  Moreover, the reduction is made in polynomial time, since all involved matrices are of polynomial size in~$l$. 

We complete Theorem~1's proof with the steps below: 
\begin{itemize}
\item Recall that
Theorem~\ref{th:inapprox_selection} shows that, unless ${\rm NP} {\in} {\rm BPTIME(\textit{n}^{\rm poly \log \textit{n}})}$, no quasi-polynomial time algorithm can distinguish between the statements S1 and S2; this implies that, under the same assumption, no quasi-polynomial time algorithm can distinguish between the statement S1$'$ and the statement~S2$'$\!.  
\item Since for any $\delta \in (0,1)$ we can take $q_1(l) =  2^{\Omega ( \log^{1-\delta} l )}$ in Theorem~\ref{th:inapprox_selection}, this implies that the smallest number of inputs rendering $[\1_{md}^\top, \0_{n-dm}^\top]$ reachable cannot be approximated within a multiplicative factor of $q_1(l)$. {Indeed, any  algorithm which gives an approximation of the smallest number of inputs with a multiplicative factor smaller than $q_1(l)$ would make it possible to distinguish between case S1$'$ and case S2$'$\!. By Theorem~\ref{th:inapprox_selection}, the inapproximability factor $q_1(l)$ grows as  $2^{\Omega ( \log^{1-\delta} l )}$\!,} and since $l$ can be upper and lower bounded by
a polynomial in $n$ (since $n\geq l$, and $n$ is at most polynomial in $l$), we set $\Delta_2(n) = 2^{\Omega( \log^{1-\delta} n )}$ in the statement of Theorem~1.
\item Since $\Delta(l)$ is a polynomial in $l$, as well as, {$l \leq n$, we may replace {$\Delta(l)$} by some polynomial $\Delta_1(n)$,} as in the statement of Theorem~1.
\end{itemize}

\section{Concluding Remarks}\label{sec:con}

We focused on the minimal reachability Problem~\ref{pr:minimal_reachability}, which is {a fundamental question in optimization and control} with applications such as power systems and neural circuits. By~exploiting the connection to the variable selection Problem~\ref{pr:selection}, we~proved that Problem~\ref{pr:minimal_reachability} is {hard to approximate}.  
Future work will focus on properties for the system matrix $A$ so that Problem~\ref{pr:minimal_reachability} is approximable in polynomial time.

{We conclude with an open problem. As we have discussed, the minimum reachability problem is $(0,2^{\log n})$-approximable by the algorithm which actuates every variable; but $(0,2^{{O} ( \log^{1-\delta} n )} )$ is impossible for any positive $\delta$.  
We wonder, therefore, whether the minimum number of actuators can be approximated to within a multiplicative factor of say, $\sqrt{n}$ in polynomial time, or, more generally, $n^c$ for some $c \in (0,1)$. }
{Indeed, observe that since $\sqrt{n} = 2^{(1/2) \log n}$\!, the function $\sqrt{n}$ does not belong to $2^{{O} ( \log^{1-\delta} n ) }$ for any $\delta>0$. Thus, the present paper does not rule out the possibility of approximating the minimum reachability problem up to a factor of $\sqrt{n}$, or more broadly, $n^c$ for $c \in (0,1)$. We remark that such an approximation guarantee would have considerable repercussions in the context of effective control, as at the moment the best polynomial-time protocol for actuation to meet a reachability goal (in terms of worst-case approximation guarantee) is to actuate every variable.}

\appendices

%
%

\ifCLASSOPTIONcaptionsoff
  \newpage
\fi



%

\bibliographystyle{IEEEtran}
\bibliography{references_reach}

%

%
%
%




\end{document}